# The milling process monitoring using 3D envelope method

BISU Claudiu[1,a], GERARD Alain[2,b], ZAPCIU Miron [1,c] and CAHUC Olivier[2,d]

[1]University Politehnica of Bucharest, Department of Machines and Systems of Production, 313 Spl. Independentei, sect.6, 060042 Bucharest.

[2]University Bordeaux 1, -I2M- Material, Processes, Interactions-CNRS, UMR 5295, 351 cours de la Liberation, 33405 Talence, France.

[a]claudiu.bisu@upb.ro , [b]ajr.gerard@gmail.com , [c]miron.zapciu@upb.ro , [d]olivier.cahuc@u-bordeaux1.fr

**Abstract**

This paper proposes a method to vibration analysis in order to on-line monitoring of milling process quality. Adapting envelope analysis to characterize the milling tool materials is an important contribution to the qualitative and quantitative characterization of milling capacity and a step by modeling the three-dimensional cutting process. An experimental protocol was designed and developed for the acquisition, processing and analyzing three-dimensional signal. The vibration envelope analysis is proposed to detect the cutting capacity of the tool with the optimization application of cutting parameters. The research is focused on Hilbert transform optimization to evaluate the dynamic behavior of the machine/ tool/workpiece.

## 1 Introduction

The whole research is to characterize the three-dimensional manufacturing system, in particular the spindle, the workpiece, to determine the imperfections or the defects of functioning due to the wear, which can modify the precision of the manufacturing. But the vibration appearance is inevitable in the dynamic cutting process particularly in the milling process. The modern CNC milling machines are widely used in modern industry for improved productivity, better precision and variety of products. Since a reduction of the production costs and an increase in the quality of the machined parts are expected, the automated detection of the machining process malfunctions has become of great interest among scientists and industrialists. Failure of cutting tools in milling significantly decreases machining productivity and quality. By the use of a large variety of sensors, monitoring of machining processes represents the prime step for reduction of poor quality and hence a reduction of costs. The dynamic monitoring analysis and replacement of the damaged tool at the right time are very important to assure machining quality and system reliability. Unfortunately, tool breakage detection in milling is difficult due to the complex nature of machining processes and the variable cutting conditions that affect the collected signals. It is desirable to develop a low cost and reliable tool breakage monitoring system for milling process **[1], [2].** Demand for better product quality and reliability has led to increased sensor integration in machine systems to enable more comprehensive, accurate, and timely gathering of information on their working status. Various sensors have been developed and employed over the past decades that measure vibration (acceleration) **[3]**, dynamic force **[4]**, acoustic emission **[5]**, or temperature **[6]** during machine operations for condition monitoring and defect diagnosis **[7]**. Since vibration signals are directly associated with the structural dynamics of the machine being monitored, vibration measurement has been widely adopted as a popular tool. Effective utilization of the vibration data, however, depends upon the effectiveness and efficiency of the signal processing technique employed to extract characteristic features (i.e. defect-induced vibration components) from the signal and assess how severe the defect in the machine system is and what needs to be done to correct the problem and ensure continuous, safe operation. This indicates that proper signal analysis is a critical prerequisite for clear identification of machine conditions, timely diagnosis of defect severity [8], and reliable prediction of the remaining service life [**9**]. This paper proposes a method of vibration analysis in order to on-line monitor the milling process quality. The method used in our research refers to an advanced analysis of vibration to obtain the answer on quality of the milling process and also to identify various defects. In order to reach to objective, an experimental device designed to obtain dynamic information provided by the dynamic system machine-tool/tool/workpiece. The main focus will be on envelope vibration analysis in order to obtain a frequency spectrum in direct connection with the quantity and the uniformity of each tooth own energy and how it is transmitted to the workpiece **[10]**.

## 2 Research scope

The main aim of the reported research is to investigate the possibility to assess the workpiece surface quality in milling by use of process monitoring. Correlation between the output signals (cutting forces, vibration) and the type of features which appeared on the workpiece surface were investigated by use of time and frequency analysis of the output signals [11]. An envelope method to milling process characterization is taken into account. The vibration signal incurred by the mill cutter is periodic impulsive signal in time domain, is a signal give by accelerator sensor. Based on the signal transfer process, the fault signal transferred into the cutter mill imposes an impulsive motivation on the mechanical system of accelerator sensor **[10]**. The purpose of this paper is to develop a method for real-time monitoring and analysis of the milling tools during the cutting process. The method used in our research refers to the spectral envelope analysis based on Hilbert transform [12] to identify mechanical defects and obtaining a better response of the milling process and also of the machine. Thus, the objective of the envelope analysis is providing real data on the milling capacity of the tool, tool wear and dynamic functionality of the assembly motor spindle with tool by emphasizing the dynamic behaviour of the bearings (figure 1).

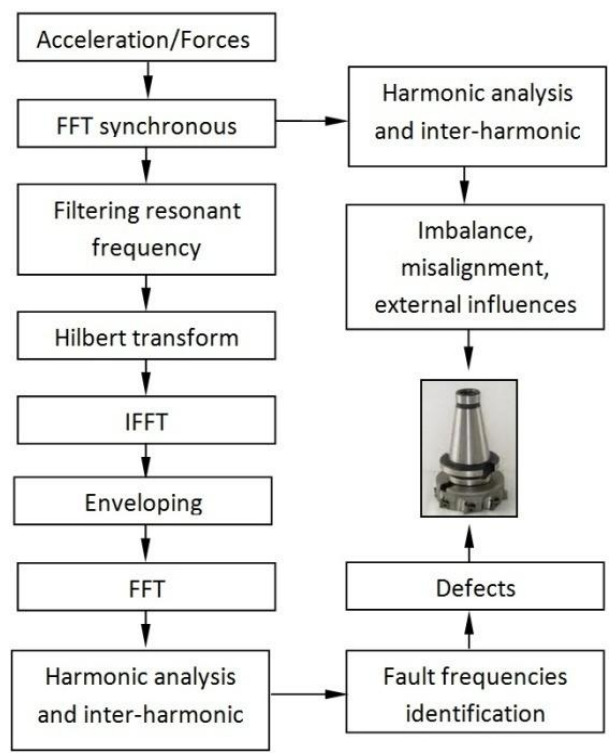

Figure 1: Envelope method.

# 3 Experimental setup

To achieve this research an experimental device is designed to obtain the dynamic information provided by the system machine/tool/workpiece. The experiments were performed on a 3 axis CNC vertical machining centre with 11 kW of power in the spindle motor and a maximum speed rotation of 8,000 rpm. Wait for our goal the recording data of vibrations and cutting forces signals in the same time with rotational speed is absolutely necessary. A Kistler 9257B stationary dynamometer Quartz 3 - Component, a National Instruments NI USB-6216 analogical/digital data acquisition board and Fastview software were used for three axis cutting force measurements, figure 2. A three-dimensional PCB piezoelectric accelerometer fixed on the workpiece and a B&K unidirectional piezoelectric accelerometer placed on the spindle in radial direction, a National Instruments NI USB-4432 analogical/digital board and Fastview software were used for vibrations measurement. The speed of rotation is achieved through a laser sensor tachometer [13]. The signals were processed with Fastview program, application developed in collaboration with Digitline Company. Before starting the dynamic analysis of the tool in cutting process, the characterization of the machine is necessary in order to identify the dynamics of the assembly tool/workpiece. An experimental protocol has been established in order to conduct a thorough analysis of the spindle in different conditions **[9]**.

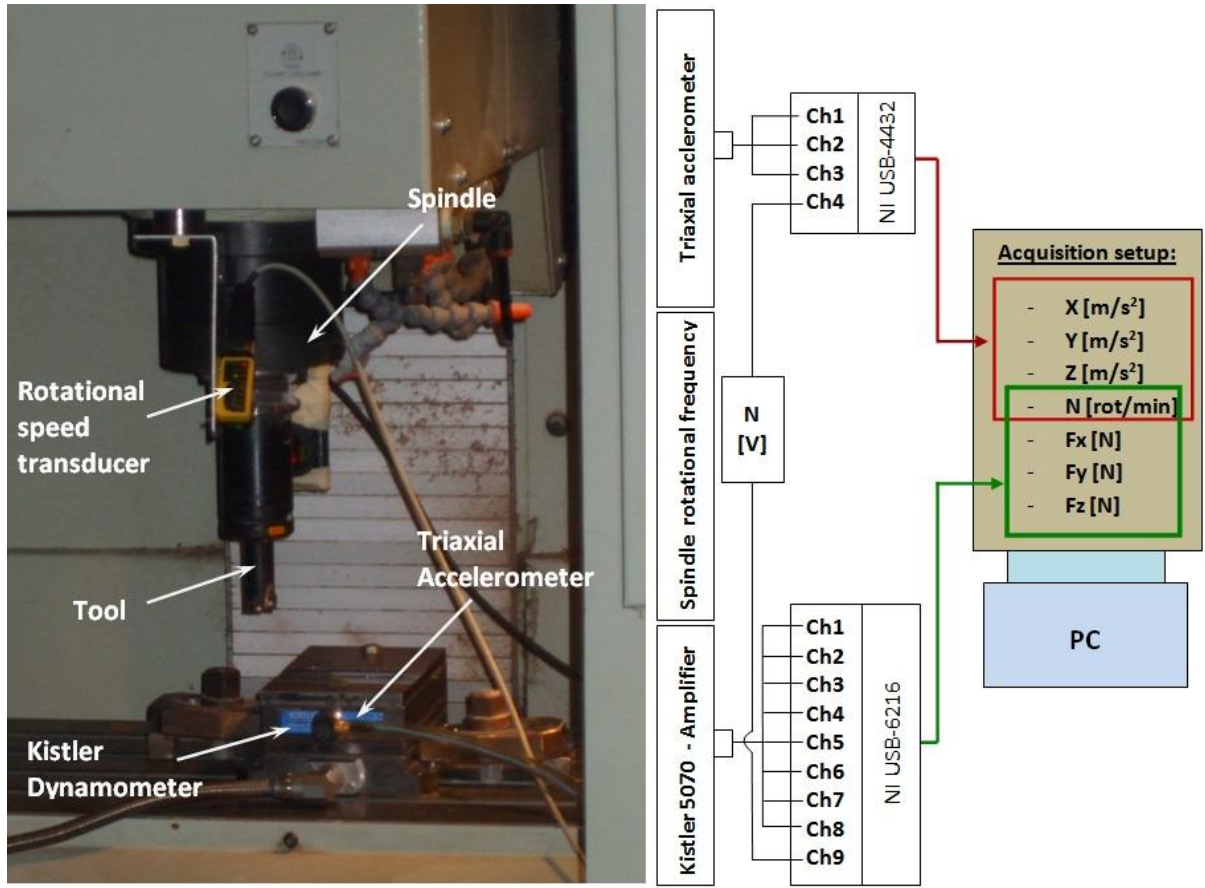

Figure 2: Experimental device.

# 4 Measurement and analysis

To increase the forces and to obtain a better response for the monitoring the test was performed on steel materials workpiece E24-2, the tool milling cutter were used here is 490-025C5-08M, with 25 mm diameter and 3 teeth. The study is focused on dynamic behaviour analysis of the mill cutter during the cutting process with 0.5 mm and 1 mm depth of cut, with 157 m/min the cutting speed and 0.1mm/tooth feed rate. In Figure 3 is presented the signals of force Fx, Fy, and Fz for cutting depth of 0.5 and 1mm for 100 mm length of cut. Characterization of dynamic milling process is performed for 1 mm depth of cut because the energy generated by the contact tool/workpiece/chip is very important compared to the depth of cut of 0.5 mm. The measurement were made with an acquisition rate of 5,000 samples/sec, a buffere size of 32,768 samples and a block size of 20,000 samples.

## 4.1 Time domain analysis

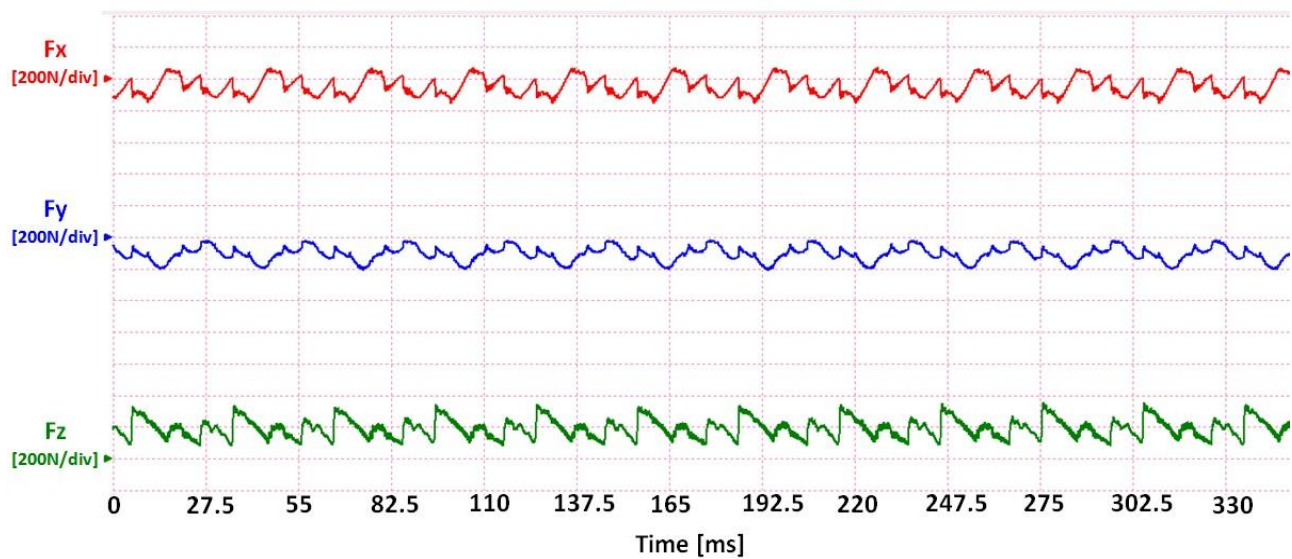

Figure 3: Waveform signal for forces of x, y and z direction for 0.5mm depth of cut.

The aim of the envelope method applied to the milling process is achieved by frequency domain processing, consistent in high accuracy synchronous FFT transform, filtering resonance band of workpiece and tool, Hilbert transform **[14]**, **[16]** followed by Inverse Fast Fourier Transform (IFFT). Next the FFT analyses of the envelope ensure high precision description of the mill cutter to identify the type and amplitude of asymmetry and wear. Each cutter tooth asymmetry is automatically qualified through the harmonic components with a lower frequency than the principal frequency equivalent of teeth number. To detect structural defects that may occur in these machine components, spectral analysis of the signal's envelope has been widely employed **[14]**, **[17]**. This is based on the consideration that structural impacts induced by a localized defect often excite one or more resonance modes of the structure and generate impulsive vibrations in a repetitive and periodic way [18].

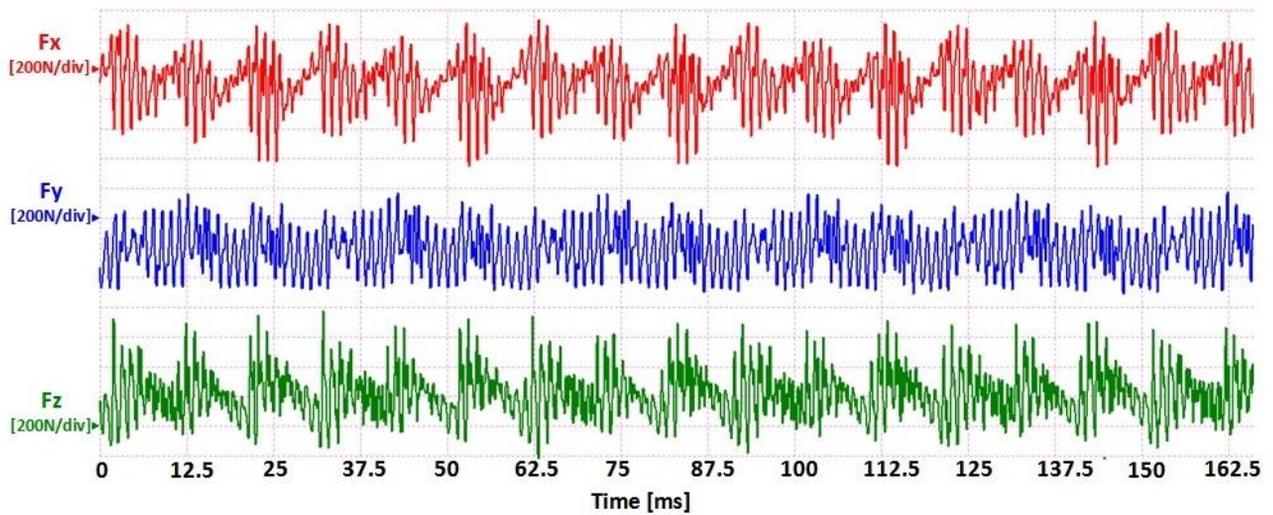

Figure 4: Waveform signal for forces of x, y and z direction for 1mm depth of cut.

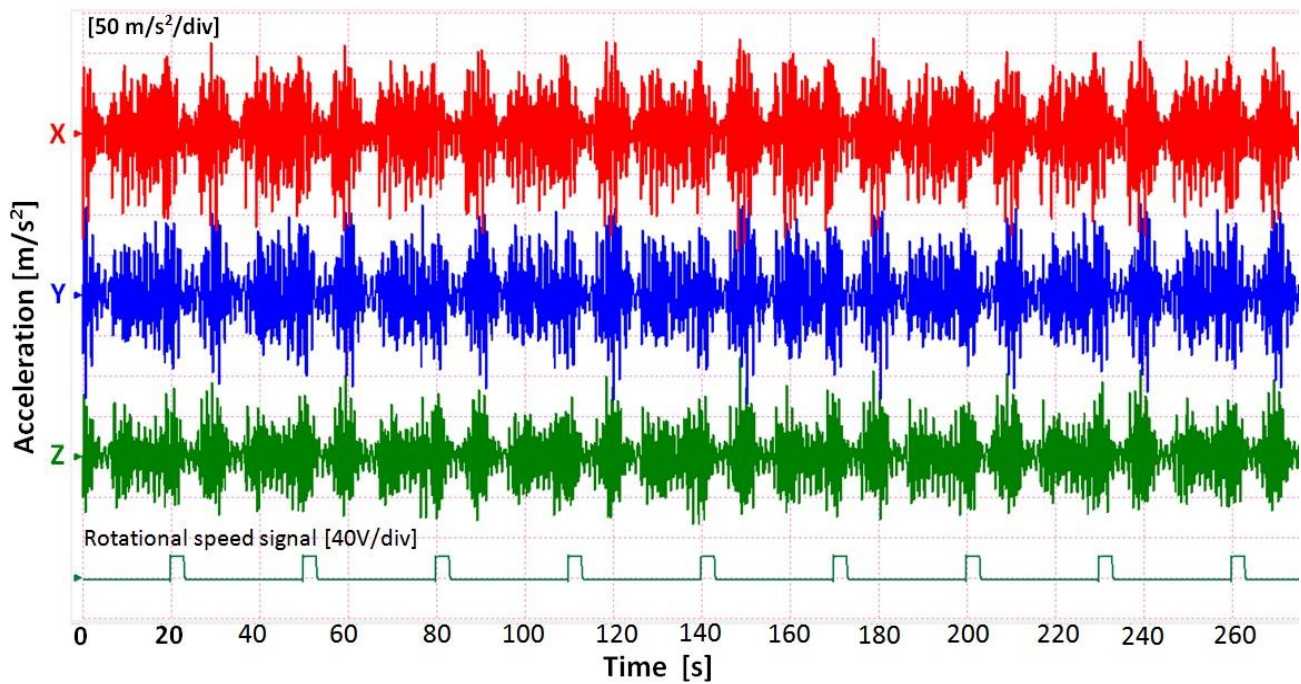

Figure 5: Waveform signal for acceleration of x, y and z direction for 1mm depth of cut.

The same evolution is observed in the case of accelerations, figure 5 shows the waveform of the acceleration signal measured on the three directions during milling processing. It can be seen that the amplitudes of the X direction (the feed direction of the tool) and Y (the cutting direction of the tool) are much higher than the Z direction (axial direction of the tool).

## 4.2 Frequency domain analysis

Forces analysis continues to apply FFT on the signal measured in time and obtain frequency spectrum due to the milling process, shows in figure 6, 7 and 8 for 1mm depth of cut.

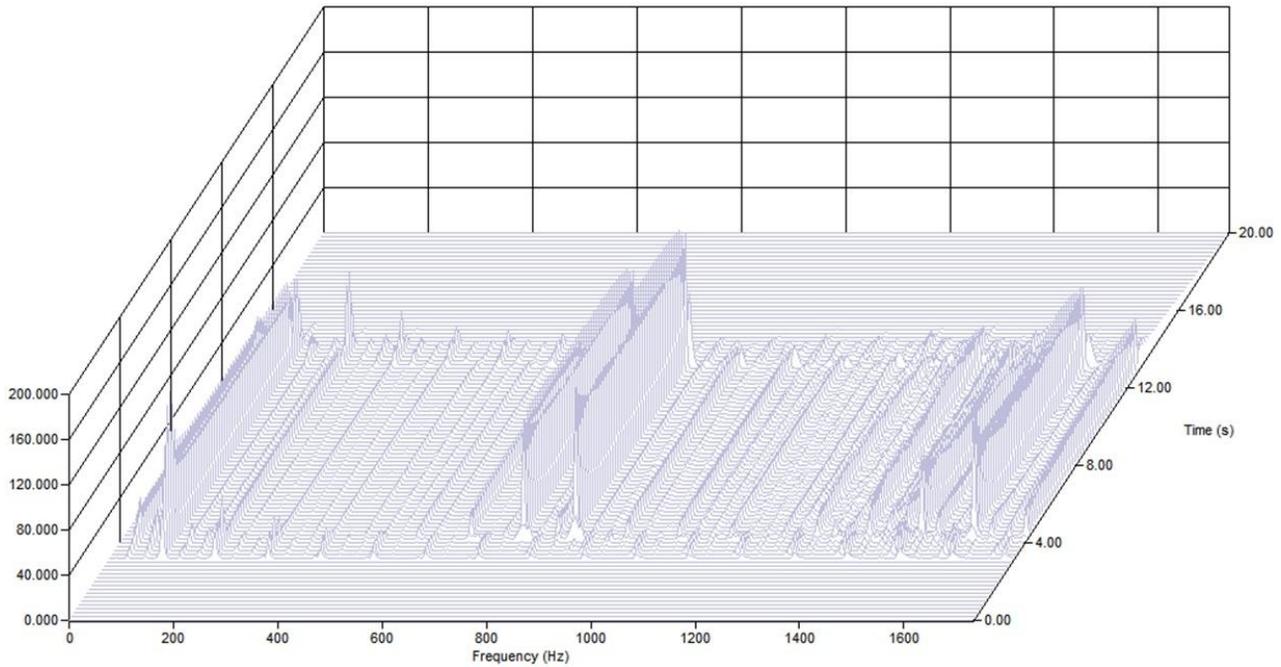

Figure 6: Waterfall diagram of Fx forces for 1mm depth of cut.

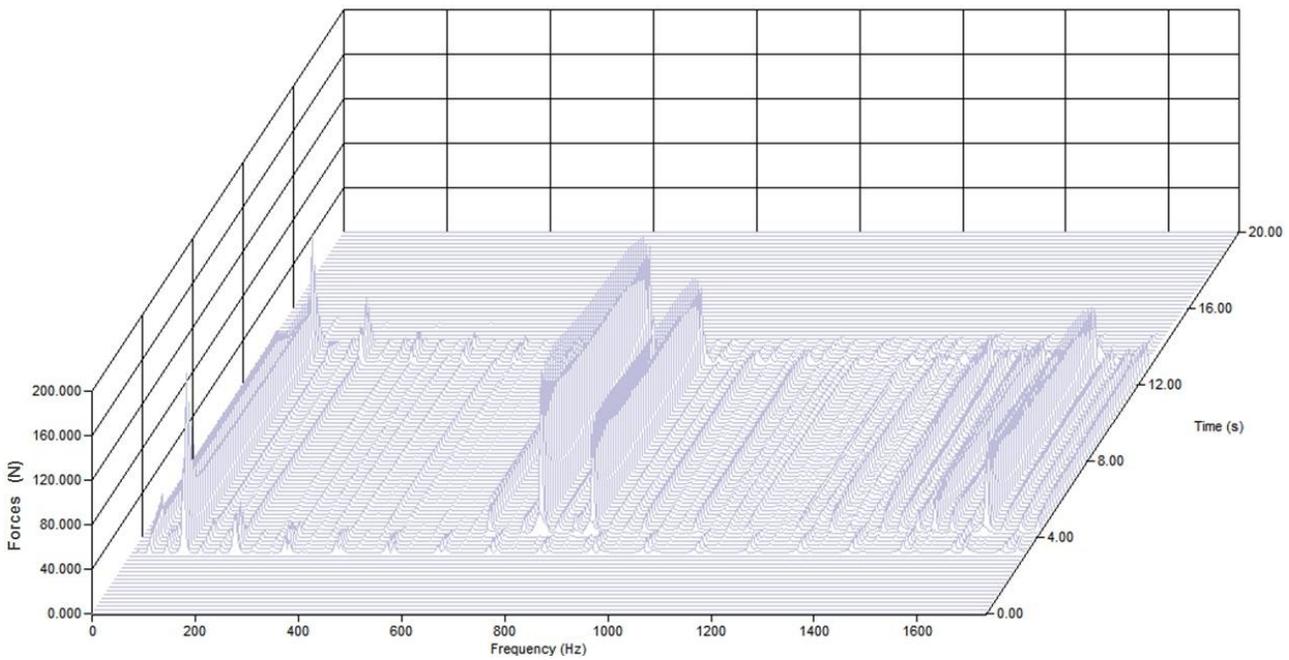

Figure 7: Waterfall diagram of Fx forces for 1mm depth of cut.

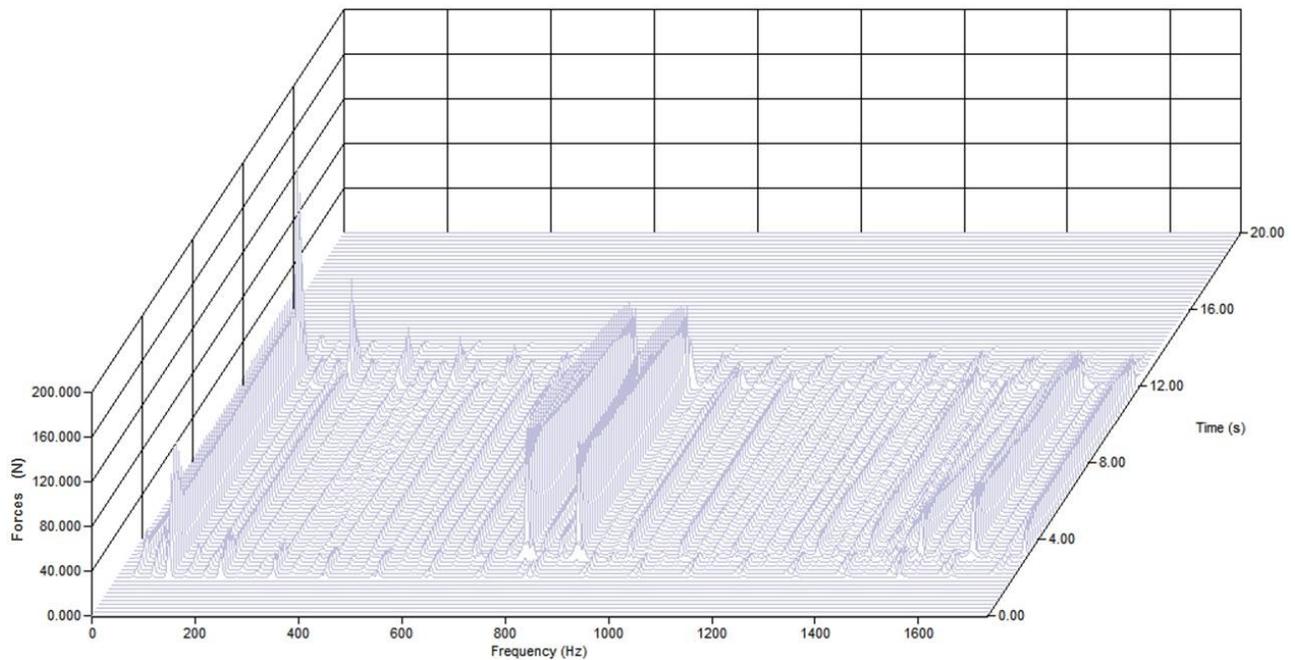

Figure 8: Waterfall diagram of Fx forces for 1mm depth of cut.

For the envelope method applications requires filtering resonance bands and then use Hilbert transform to find the periodic data impacts of cutter teeth. The resonance band for filtering is 700-2, 500 Hz.

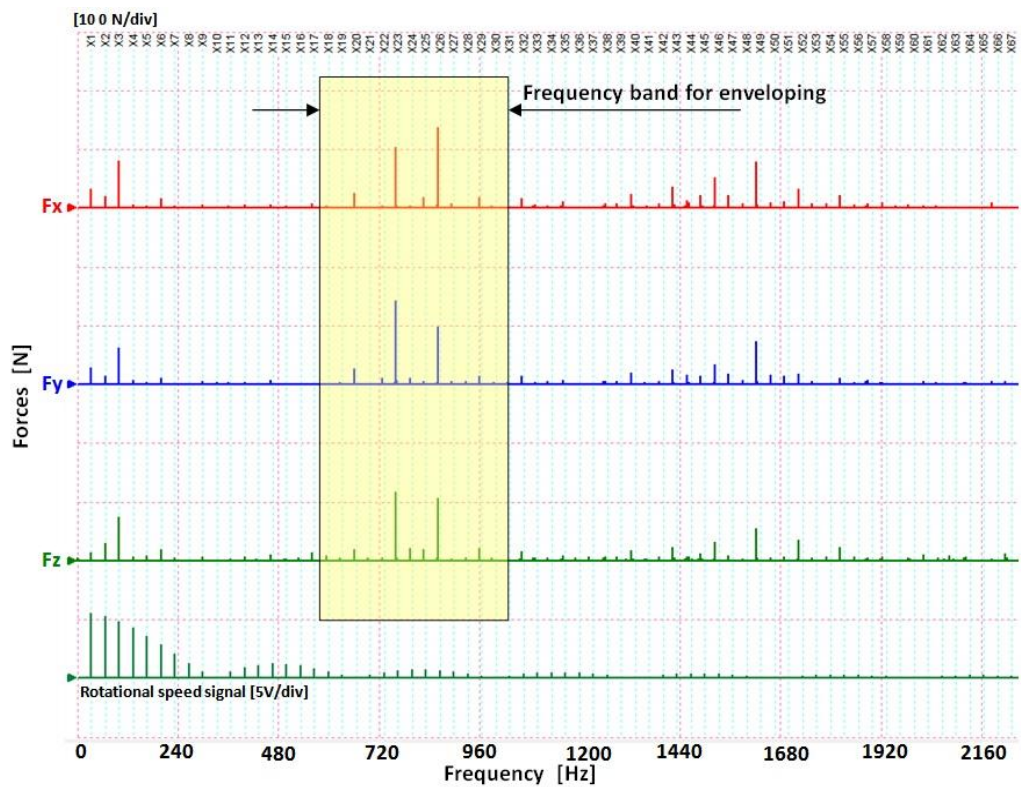

Figure 9: FFT forces signal for 1mm depth of cut and the filter band for enveloping method.

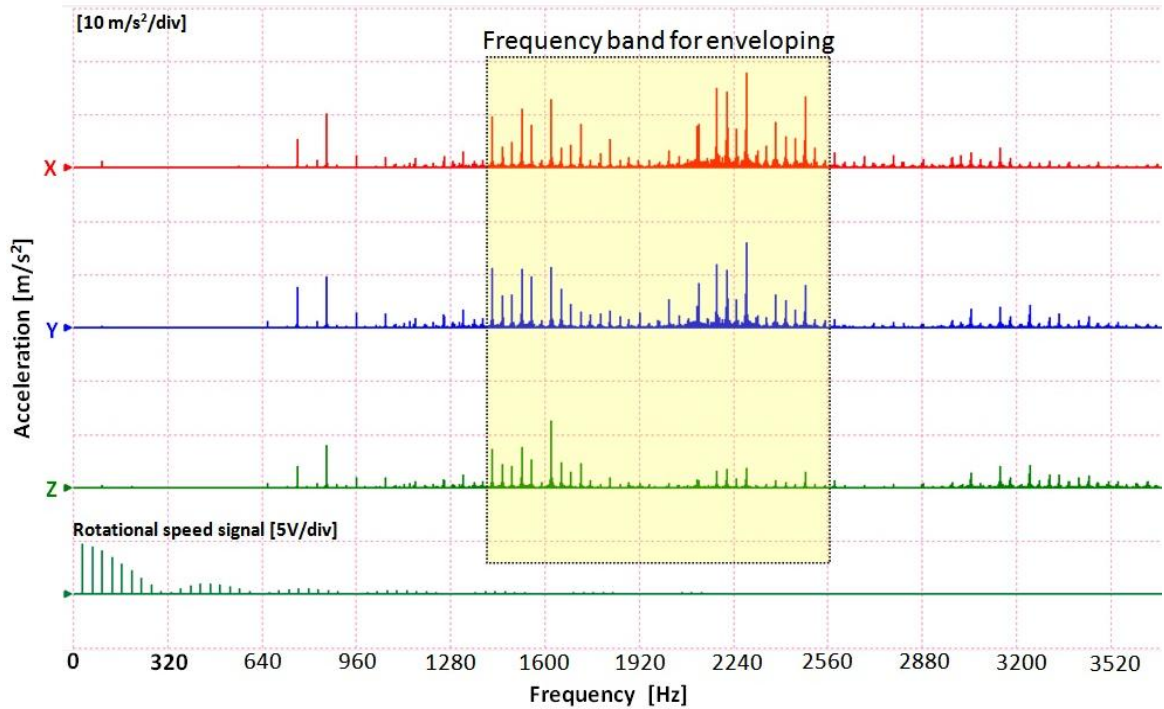

Figure 10: FFT vibration signal for 1mm depth of cut and the filter band for enveloping method.

### 4.3 Envelope analysis

Frequencies related to such resonance modes are often located in higher frequency regions than those caused by machine-borne vibrations, and are characterized by an energy concentration within a relatively narrow band centred at one of the harmonics of the resonance frequency.

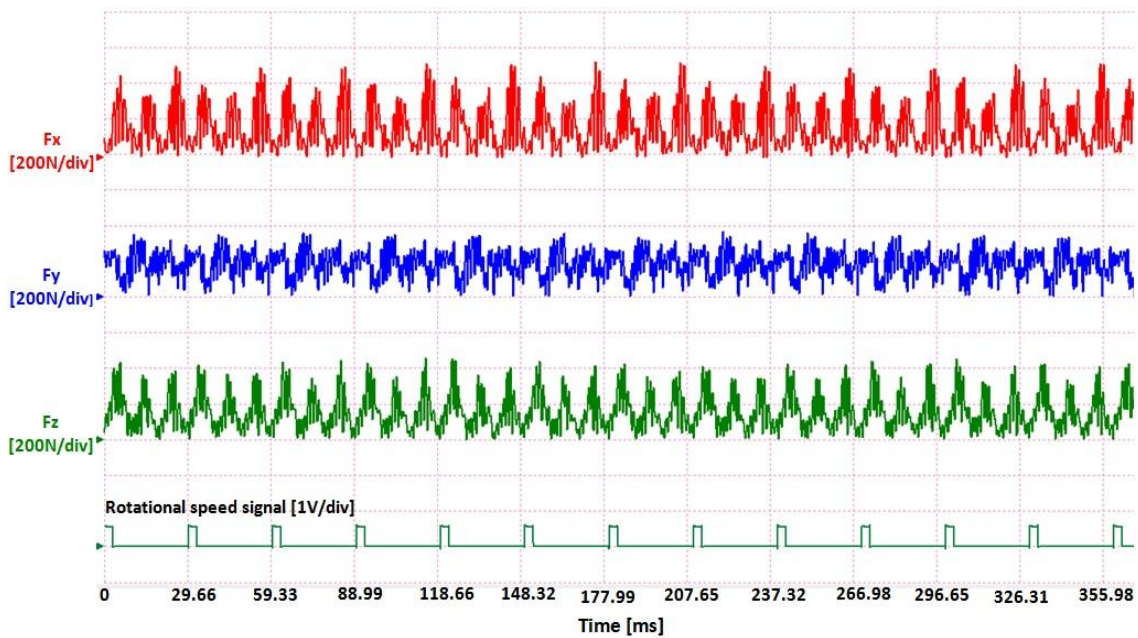

Figure 11: The waveform signal enveloping for forces.

By utilizing the effect of mechanical amplification provided by structural resonances, defect-induced vibration features can be separated from the background noise and interference for diagnosis purpose **[14].**

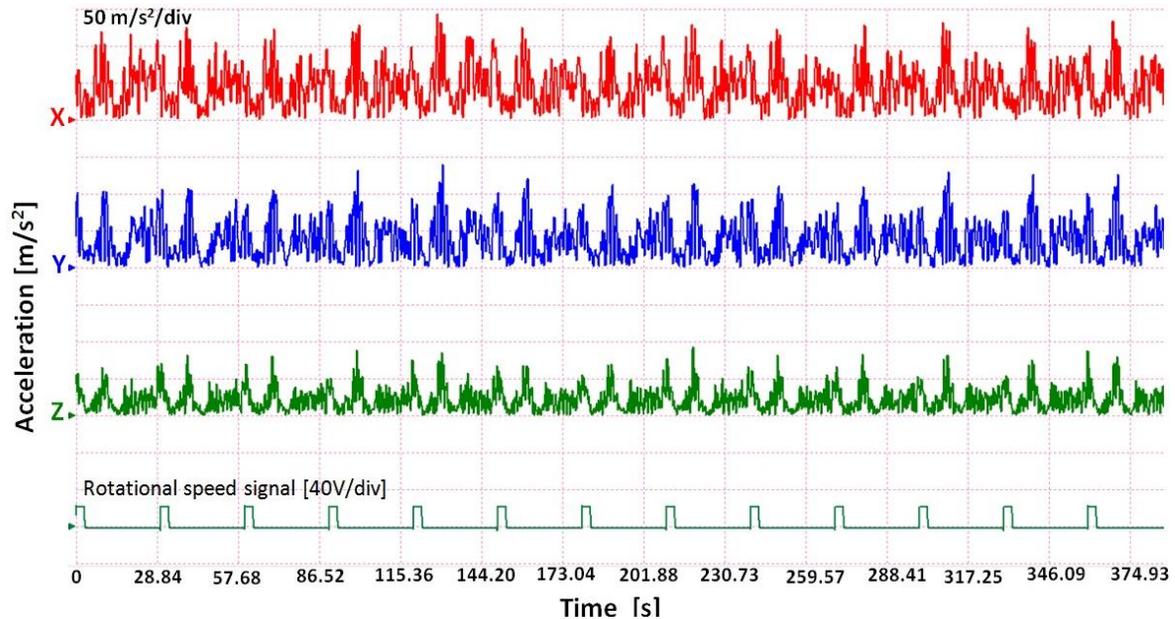

Figure 12: The waveform signal enveloping for vibration.

With dynamic information tool and workpiece can apply the enveloping synchronous method to evaluate with high precision the cutting quality of the tool (figure 9 and figure 10). The objective here is to accurately track the transfer power by tooth workpiece contact and its harmonic distribution. The aim of the envelope method applied to the milling process is achieved by frequency domain processing, consistent in high accuracy synchronous FFT transform, filtering resonance band of workpiece and tool, Hilbert transform **[17]** followed by Inverse Fast Fourier Trans- form (IFFT), figure 11 and figure 12. Next the FFT analyses of the envelope ensure high precision description of the mill cutter to identify the type and amplitude of asymmetry and wear (figure 13 and figure 14). Each cutter tooth asymmetry is automatically qualified through the harmonic components with a lower frequency than the principal frequency equivalent of teeth number. By applying FFT on the signal envelope can be observed harmonic frequency of order 3 corresponding to the activity of the three teeth but very important is the existence of order 2, were the amplitude is very close to the 3nd order harmonic. This 2nd order harmonic shows the asymmetry existence of the tool cutter. This effect leads to a cause of wear of the teeth cutter or even more to misalignment teeth/tool cutter.

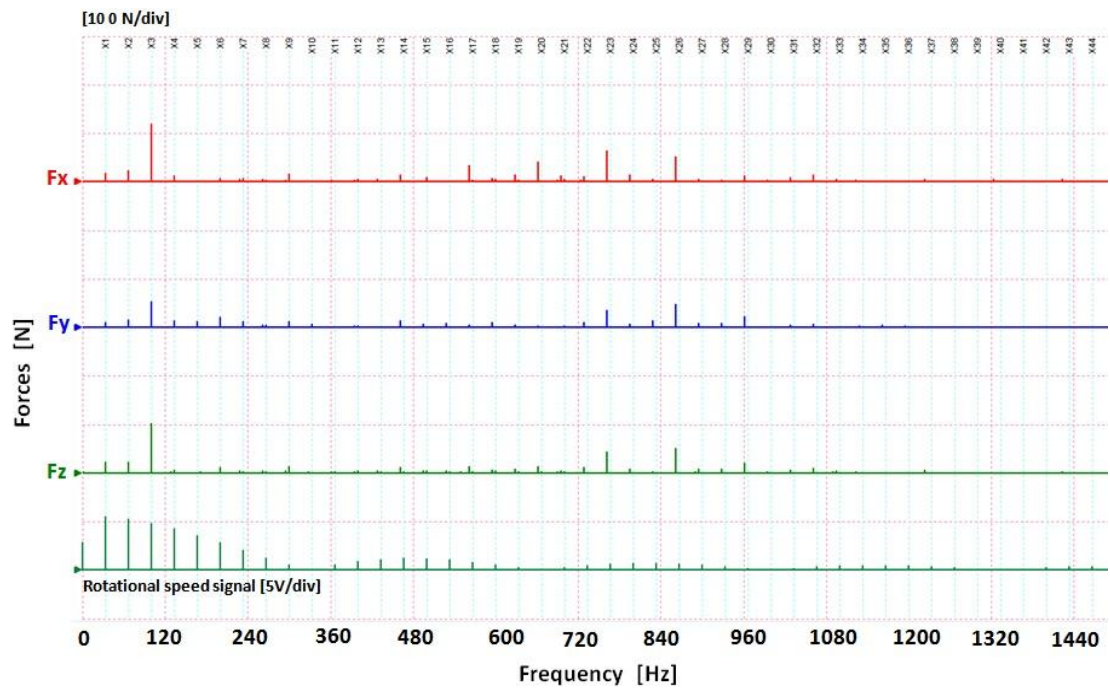

Figure 13: FFT signal enveloping for 1mm depth of cut in the forces case.

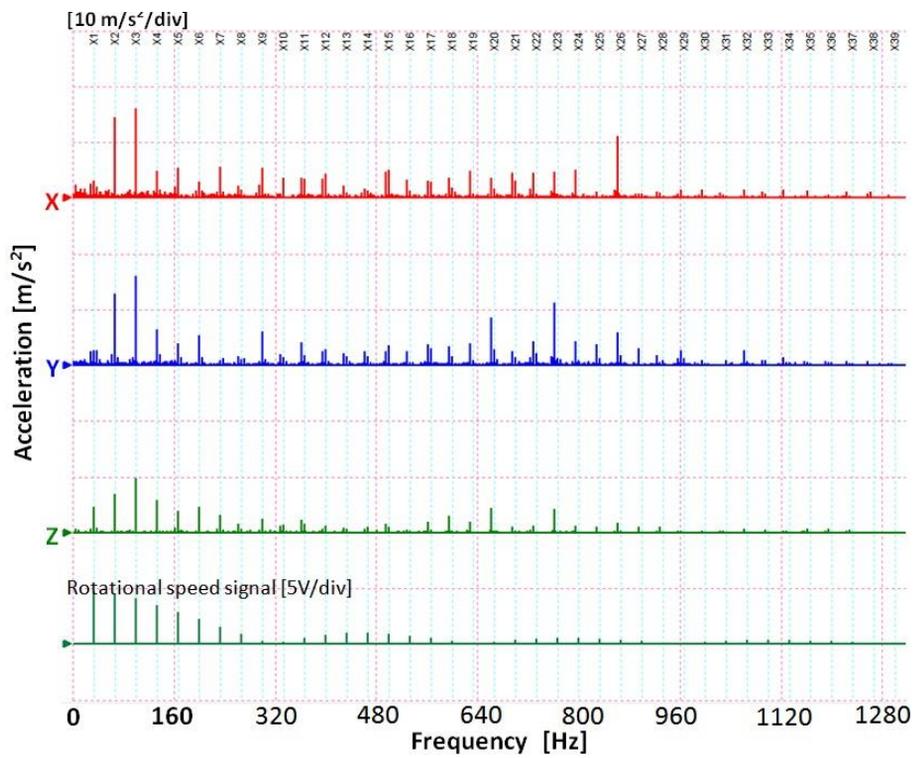

Figure 14: FFT signal enveloping for 1mm depth of cut in the vibrations case.

# 5 Conclusion

Research on vibration analysis of rotating elements has developed surveillance techniques by the methods of envelope to detect defects on bearings or gears **[19], [20]**. The transposition of these methods adapted to the field of machining combined with sampling techniques and signal processing by applying FFT synchronous and Hilbert transform has demonstrated a very promising results. This paper proposed a method of dynamic analysis based on envelope analysis method with the purpose to identify and evaluate the dynamic behaviour of the tool during cutting process. An experimental protocol was designed and developed for the acquisition, processing and analyzing of the three-dimensional vibration and force signal. The vibration signals are the result of a mixture of different sources corresponds to components of machines, making it difficult to identify the state of damage to a particular component. Adapting envelope spectral analysis to characterize the milling tool is an important contribution for the qualitative and quantitative characterization of milling capacity. The vibration envelope analysis is proposed to detect the cutting capacity of the tool necessary for process quality on-line monitoring. In under way this method represents a source for cutting parameters optimization. Is useful both, dynamic characterization of the tool and also for the monitoring process and maintenance. If the vibration acceleration signals are the source of chatter vibration, it would develop the indicators able to detect one of the most problematic phenomena in machining. In the future we are interested in creating a dynamic three-dimensional model **[21], [22**

**Acknowledgement**:This paper was supported by CNCSIS-UEFISCSU, project PNII-RUcode194/2010.